\documentclass[usenatbib]{basi}
\usepackage[T1]{fontenc}
\usepackage[british]{babel}
\usepackage[varg]{txfonts}
\usepackage{rotating}
\usepackage{dcolumn}

\usepackage{xcolor}

\newcommand\aj{{AJ}}
\newcommand\apj{{ApJ}}
\newcommand\apjs{{ApJS}}
\newcommand\aap{{A\&A}}
\newcommand\pasp{{PASP}}
\newcommand\nat{{Nature}}
\newcommand\procspie{{Proc.~SPIE}}

\begin{document}
\title[SPLAT]{The SpeX Prism Library Analysis Toolkit (SPLAT): A Data Curation Model}
\author[A.~J.~Burgasser]%
       {A.~J.~Burgasser$^1$\thanks{email: \texttt{aburgasser@ucsd.edu}} 
       and the SPLAT Development Team\thanks{This proceedings describes work developed through contributions by C.~Aganze, D.~Bardalez Gagliuffi, J.~Birky, C.~Choban, A.~Davis, I.~Escala, A.~Iyer,       Y.~Jin,       M.~Lopez,      G.~Mercado,       E.~H.~ Moreno,       J.~Parra,       M.~Sahi,       G.~Shippee       A.~Suarez,       M.~Tallis,       C.~Theissen,       T.~Tamiya, and   R.~van Linge}\\
       $^1$UC San Diego, M/C 0424, 9500 Gilman Drive, La Jolla, CA 92093, USA\\
       }
\pubyear{2017}
\volume{00}
\pagerange{\pageref{firstpage}--\pageref{lastpage}}

\date{Received --- ; accepted ---}

\maketitle
\label{firstpage}

\begin{abstract}
%
I describe our team's development of the SpeX Prism Library Analysis Toolkit (SPLAT), a combined spectral data repository for over 2500 low-resolution spectra of very low mass dwarfs (late M, L and T dwarfs), and Python-based analysis toolkit. SPLAT was constructed through a collaborative, student-centered, research-based model with high school, undergraduate and graduate students and regional K-12 science teachers. The toolkit enables spectral index analysis, classification, spectrophotometry, atmosphere model comparison, population synthesis, and other analyses. I summarize the current components of this code, sample applications, and future development plans. 
\end{abstract}

\begin{keywords}
   astronomical data bases: miscellaneous --- stars: low-mass, brown dwarfs --- techniques: spectroscopic

\end{keywords}





\section{Introduction}\label{s:intro}

The spectra of the lowest-mass stars and brown dwarfs, collectively {known as}  ``ultracool'' dwarfs (late M, L, T, and Y spectral classes),
are rich in atomic and molecular diagnostics of effective temperature, surface gravity, {chemical} composition, cloud properties, atmospheric dynamics, and multiplicity. Several libraries of ultracool
dwarf optical and infrared spectra are available online (e.g., IRTF Spectral Library, \citealt{2009ApJS..185..289R}; NIRSPEC Brown Dwarf Spectroscopy Survey; \citealt{2003ApJ...596..561M}), and have proven useful
for characterizing brown dwarf and exoplanet atmospheres, testing low temperature
atmospheric models, and designing filters and photometric search criteria, among other applications.
However, these archives do not provide tools to conduct 
standard analyses of the data, such as spectrophotometry or classification, requiring users to invest time and resources to develop these tools, and limiting the participation of novice researchers (e.g., students, citizen scientists).  Collaborative code development in astronomy, exemplified by the \texttt{astropy} python library \citep{2013A&A...558A..33A}, \texttt{dotastronomy} community,
and collaborative ``hack day'' sessions, demonstrate a means of developing community research
tools tied to data repositories, transforming them into data curations.

In this contribution, we describe the SpeX Prism Library Analysis Toolkit (SPLAT), a data curation that combines a large collection of ultracool dwarf spectra with python-based analysis tools. In the following sections, I describe the development of the code, highlight key functionality, and identify paths for future development.  The SPLAT code can be accessed at \texttt{https://github.com/aburgasser/splat/}.
 
\section{Development of SPLAT}\label{s:dev}

SPLAT grew from the SpeX Prism Libraries (SPL; \citealt{2014ASInC..11....7B}), on online repository of  nearly 1000 low-resolution ($\lambda/\Delta\lambda$ $\approx$ 120), near-infrared (0.8--2.4~$\mu$m) spectra of ultracool dwarfs obtained by the community using the NASA Infrared Telescope Facility (IRTF) SpeX spectrograph \citep{2003PASP..115..362R}. As a large uniform dataset of ultracool dwarfs, the SPL
has been used in over 100 studies of stars, brown dwarfs, exoplanets, and high redshift galaxies. The SPL  enables users to download individual and sets of spectra, and provides basic visualization; but it is a static repository with limited source information and no tools to analyze the data files.

\begin{figure}[t]
  \centerline{\includegraphics[height=3.7cm]{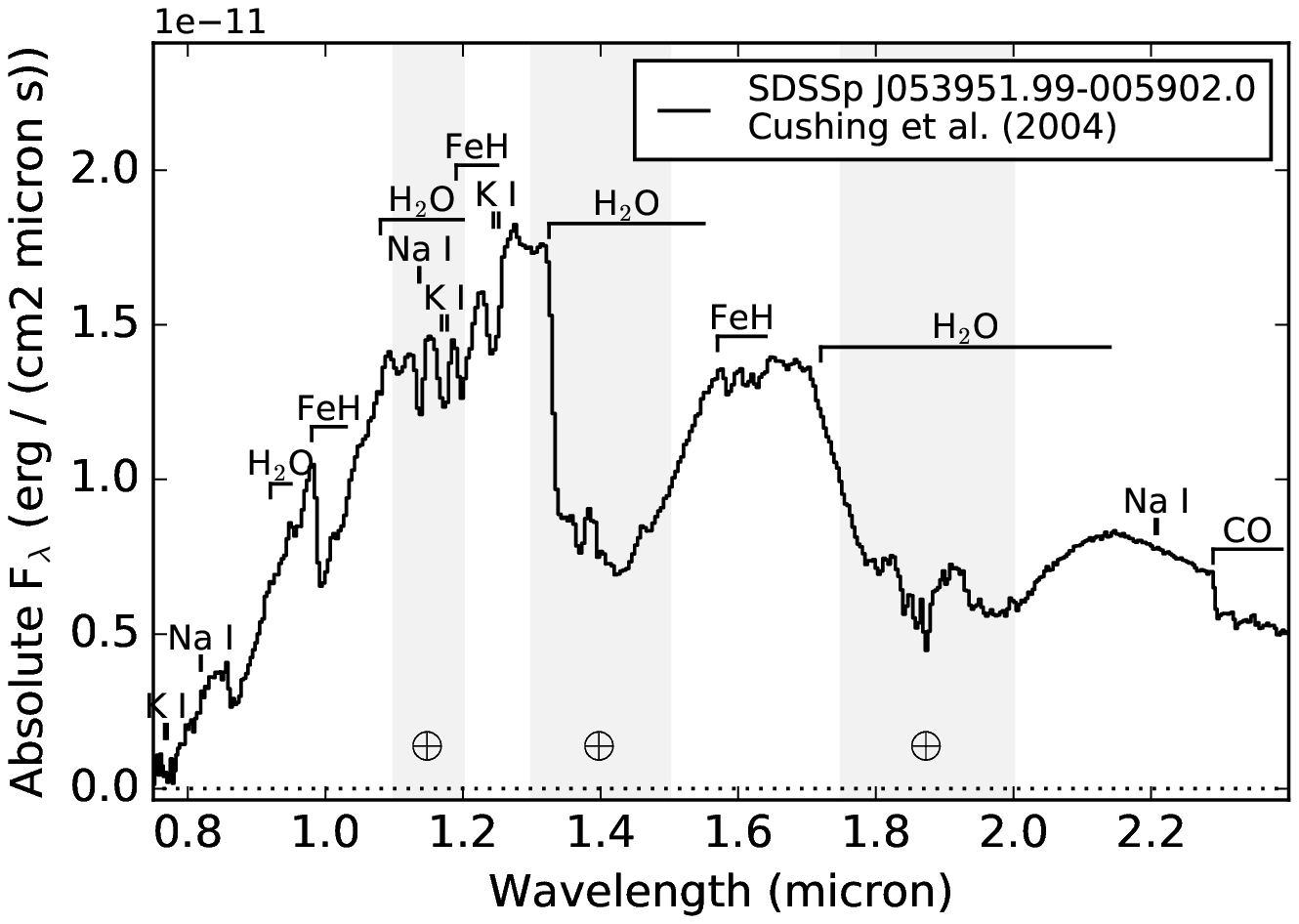}
  \includegraphics[height=3.7cm]{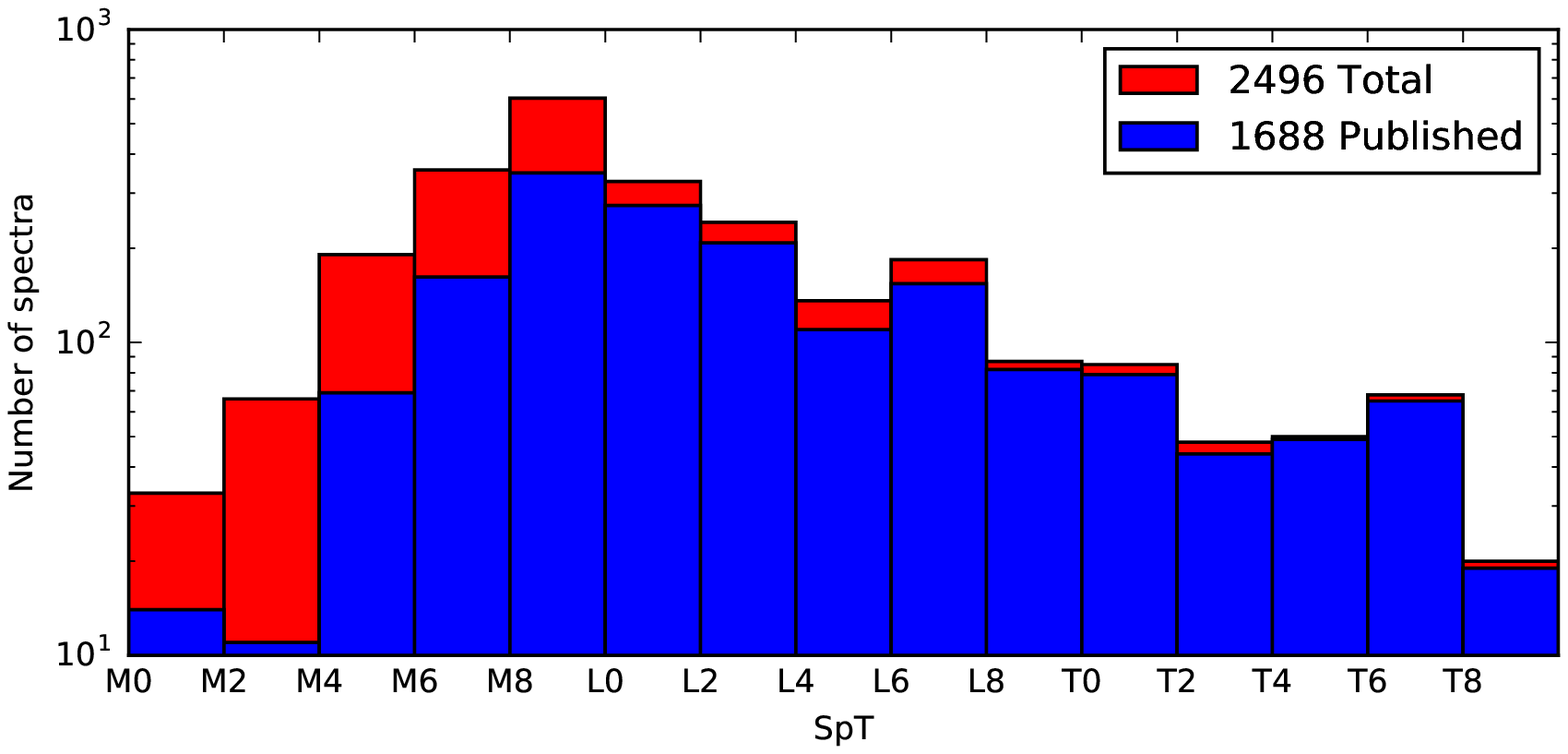}}
  \caption{(Left): Example L dwarf IRTF/SpeX spectrum from SPLAT, with features labeled. (Right): Histogram of the M, L, and T dwarfs with spectra in SPLAT, separating total (2496 spectra) and published data (1688 spectra). There are $\sim$300 spectra of other astronomical objects in the Library.}
  \end{figure}

Starting in 2013, our group began development of a Python-based analysis package for these data, partly to convert existing code from IDL to Python, and partly to engage students in programming and software development. The development process was designed to be collaborative, student-centered, and research-based. Analysis tools were generated through a series of quarterly exercises, steadily increasing in complexity; e.g., from ``read in a spectrum and plot it'' to ``perform a Markov Chain Monte Carlo {(MCMC)} fit of a spectrum using three sets of models''. These tools were built using standard Python libraries (e.g., \texttt{numpy}, \texttt{scipy}), \texttt{astropy} routines, and other community software (e.g., \texttt{emcee}; \citealt{2013PASP..125..306F}). Concurrent student-led research refined the tools through practical applications. The spectral library was also expanded to more than twice the size of SPL with new observations and contributions from the community. Source contextual data (e.g., photometry, astrometry, citations) were added, and theoretical atmosphere and evolutionary models were integrated to facilitate analysis of the data.

As of April 2017, the library contains roughly 2500 ultracool dwarf spectra, encompassing over 90\% of known late-M, L and T  dwarfs viewable from IRTF ($-50^o < \delta < +67^o$) and within 25~pc of the Sun (Figure~1). It also contains a smaller number ($\sim$300) of other sources, including planetary bodies (e.g., Saturn), galaxies, giant stars, white dwarfs, carbon stars, and supernovae.

\section{SPLAT Tools and Components}\label{s:tools}

\begin{table}[t]
  \caption{Modules in SPLAT.}\label{tab:module}
  \medskip
  \begin{center}
    \begin{tabular}{ll}
    \hline
    \multicolumn{1}{c}{Module}
                 & \multicolumn{1}{c}{Functions} \\
     \hline
core.py & Spectrum object; combination \& comparison; scaling, smoothing \& reddening; \\
 & indices \& equivalent widths; classification \\
plot.py & Visualization, including batch plotting of multiple spectra \\
photometry.py & Spectrophotometry tools and flux/magnitude conversion \\
empirical.py & Conversions between spectral type and absolute magnitude, $T_{eff}$, luminosity, \\
 & \& color; distance estimation \\
model.py & Interface for theoretical spectral models \& model-fitting routines \\
evolve.py & Interface for evolutionary models and population synthesis tools \\
database.py & Database access and import tools; SIMBAD \& Vizier through \texttt{astroquery} \\
citations.py & Bibliographic tools linked to \texttt{BibTeX} \& NASA ADS \\
utilities.py & Program checks; conversions for coordinates, date information, \& spectral  \\
 & types; basic statistical tools \\
initialize.py & Program constants \\
    \hline
    \end{tabular}\\[5pt]
  \end{center}
\end{table}

SPLAT tools are organized into modules based on categories of analysis. The spectral data themselves are manipulated through SPLAT's Spectrum class, a container for both the spectral data and associated contextual information. This class also provides baseline functions, including math operations (with robust noise propogation); wavelength and flux conversion (including $F_{\nu} \leftrightarrow F_{\lambda}$, brightness temperature, and radial velocity shifting); smoothing, scaling and reddening; and exporting of data to files. Users can work with data pulled from the SPLAT library or upload their own spectral files.  SPLAT can also``stitch'' spectra covering different wavelength ranges.

Core spectral analysis tools include spectral index measurement for user-defined indices or index sets from the literature, equivalent width measurement, and comparison of spectra, all with Monte Carlo noise propagation and a variety of masking options. These base tools are utilized in SPLAT's classification suite, which type ultracool dwarfs using indices, comparison to M0-T8 standards defined in \citet{2010ApJS..190..100K}, and comparison to the entire library of spectra.  There is also an implementation of the \citet{2013ApJ...772...79A} index-based scheme for surface gravity classification.
SPLAT's visualization suite is designed for near-infrared spectral data, with preset feature labeling, masking of telluric bands, and visual comparison of spectra (e.g., for standard classification). There are batch plotting options for sets of spectra.
SPLAT's spectrophotometric suite contains over 100 standard (e.g., SDSS, 2MASS, MKO) and instrument-defined filters (e.g., HST/WFC3, Keck/NIRC2, Magellan/FourStar), and users can measure magnitudes, fluxes, and photon counts on AB and Vega magnitude systems; or flux calibrate spectra to observed apparent or absolute magnitudes.

With its focus on ultracool dwarf science, SPLAT routines provide parameter conversions between spectral type and absolute magnitude, effective temperature ($T_{eff}$), luminosity, and color using published empirical relations. There is also a dedicated spectrophotometric distance estimator.
SPLAT has tools for fitting spectra to theoretical atmosphere models, drawn from \citet{2012RSPTA.370.2765A,2006ApJ...640.1063B,2011ApJ...737...34M,2011A&A...529A..44W,2012ApJ...756..172M,2014ApJ...787...78M} and \citet{2012ApJ...750...74S}. The models, generally spanning 500~K $\lesssim$ $T_{eff}$ $\lesssim$ 3000~K and with various condensate cloud treatments, are smoothed and constrained in wavelength to match the SpeX data. Low-resolution ($\lambda/\Delta\lambda$ = 100), broad-band (0.3-30~$\mu$m) model spectra are also available for spectral energy distribution analysis. SPLAT provides three routines for fitting models to data using a variety of statistics, with comparisons made to both model grid points and log-linear interpolated models. There are two {MCMC} modeling fitting routines: a native implementation of the Metropolis-Hastings algorithm and the {\it emcee} implementation of the affine-invariant MCMC \citet{2013PASP..125..306F}.
SPLAT also contains theoretical evolutionary models for ultracool dwarfs from \citet{2001RvMP...73..719B,2003A&A...402..701B} and \citet{2008ApJ...689.1327S}, which allow users to transform between empirical measurements of temperature, surface gravity or luminosity to physical parameters of mass, radius and age.  This module also contains synthetic sample generation routines for population synthesis analysis.

SPLAT's database routines are primarily designed for access to the internal spectral and source databases, including library searches and importing tools. Through the astroquery package \citep{adam_ginsburg_2016_44961}, SPLAT provides access to Vizier and SIMBAD so users can acquire contextual data. 
Finally, SPLAT contains a number of utility tools that enable manipulation of bibliographic information (through BibTeX and NASA ADS queries), conversion routines for coordinates and dates, and some basic statistical tools.

\section{SPLAT in Action}\label{s:action}

\begin{figure}[t]
  \centerline{  
  \includegraphics[height=4cm]{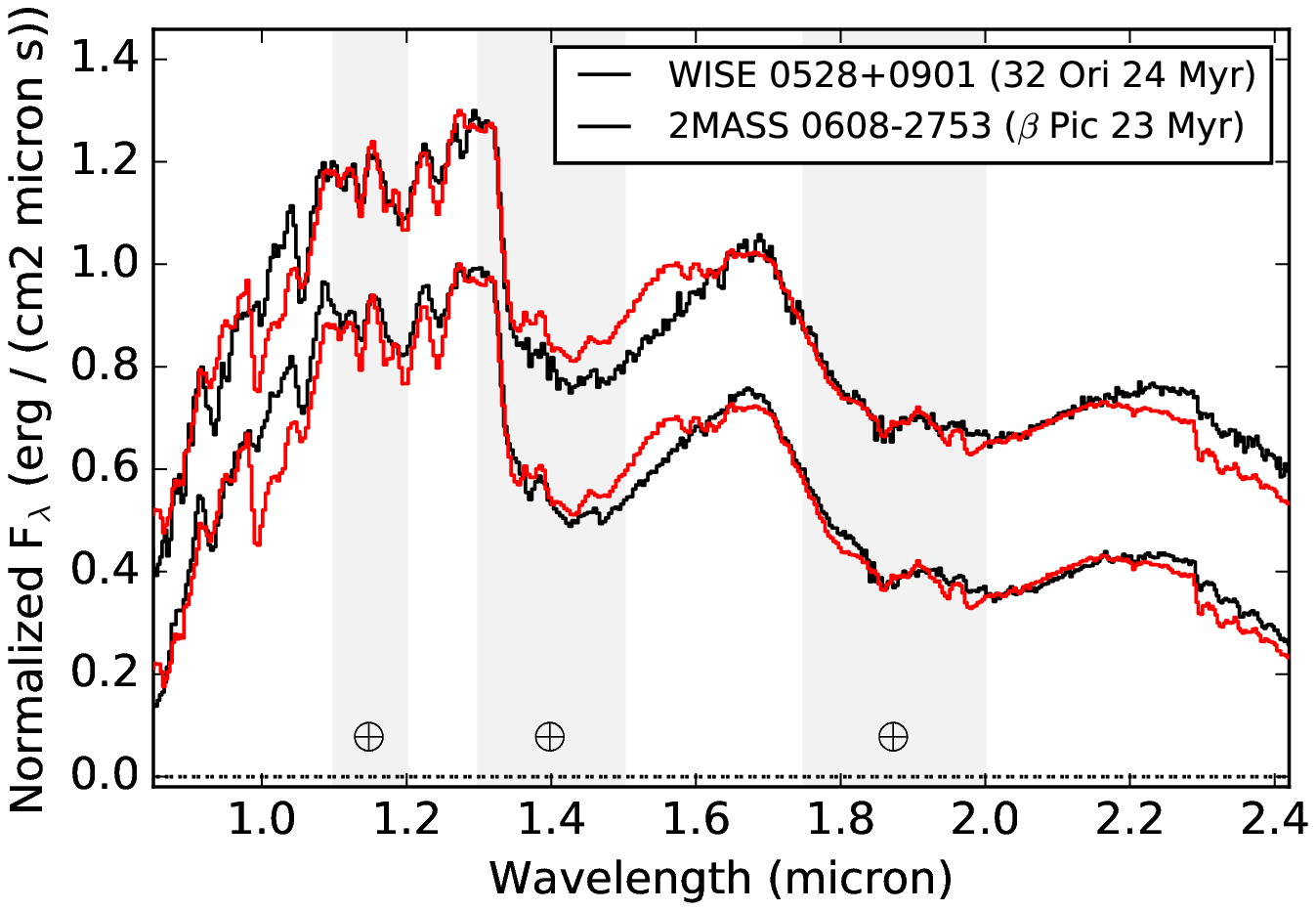}
  \includegraphics[height=4cm]{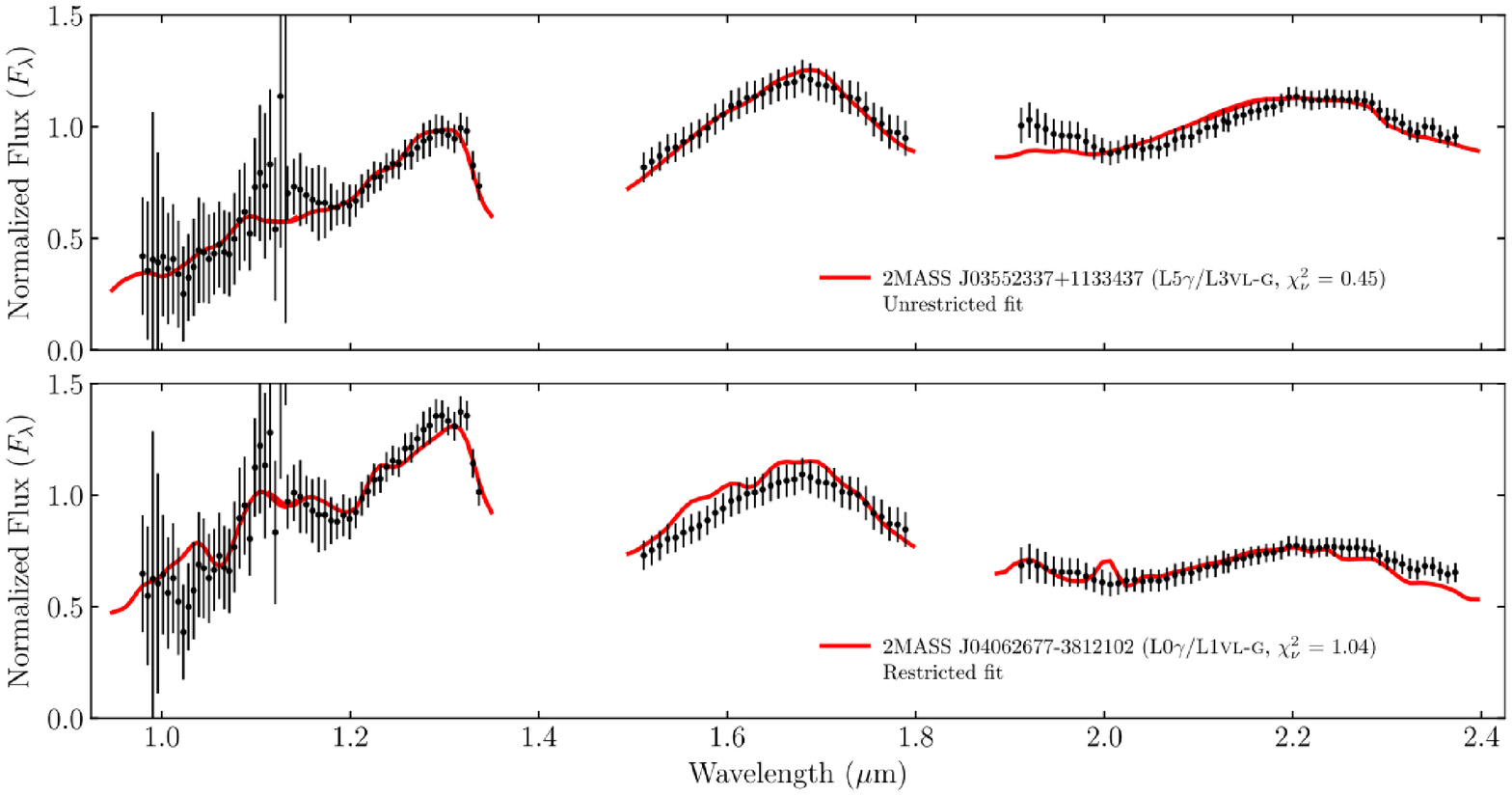}
  }
  \caption{Examples of SPLAT in action: 
  (Left) Spectrum of the 32 Orionis planetary-mass brown dwarf WISE~J0528+0901 (top) compared to a comparably-aged member of the $\beta$ Pictoris association, 2MASS~J0608$-$2753 (bottom), and the L1 spectral standard 2MASS~J2130$-$0845 (red overlaying both spectra).
  (Right) Comparison of SPLAT templates (red) 2MASS~J0355+1133 (top) and 2MASS~J0406-4812 (bottom) to the GPI spectrum of $\beta$ Pictoris b (black points with error bars; from \citealt{2017AJ....153..182C}).
  \label{fig:examples}}
  \end{figure}

SPLAT has already been used in a number of studies published in the literature examining the spectra of cool brown dwarfs and exoplanets (Figure~\ref{fig:examples}).
In \citet{2016ApJ...820...32B}, SPLAT tools were used to identify the L dwarf WISE~J0528+0901 as a young planetry-mass brown dwarf, whose kinematics place it in the 24~Myr-old 32 Orionis association.
In \citet{2016AJ....151...46A}, SPLAT tools were used to both classify and measure the metallicity of the very low-mass secondary in the GJ~660.1 system through spectral model fits.
In \citet{2016Natur.533..221G}, SPLAT tools were used to determine the metallicity of the planet-host star TRAPPIST-1 using equivalent width measurements and empirical calibrations.
\citet{2016ApJ...824..121D,2017AJ....153..182C}, and \citet{2017AJ....153..190J}, SPLAT tools were used the characterize Gemini Planet Imager (GPI; \citealt{2014SPIE.9148E..0JM}) spectra of the directly-imaged exoplanets HD 95086b, $\beta$ Pictoris b and HD 984B, respectively

\section{Future Directions}\label{s:future}

With fewer than 3000 spectra and an equivalent number of models, SPLAT is not a ``big data'' resource, so there is interest in expanding the library and toolkit for use with data from other infrared spectrographs (e.g., SpeX cross-dispersed mode, APOGEE, Magellan/FIRE, VLT/X-shooter, various/TripleSpec) and from optical spectrographs (e.g., SDSS). While we have demonstrated SPLAT's use for high resolution data  \citep{2016Natur.533..221G}, a broader expansion will require generalizing the Spectrum class and many of the analysis routines to these other data formats.  Replacement of the Spectrum class is likely imminent, as \texttt{astropy} has recently begun the process of integrating spectral analysis tools into its basecode (\texttt{specutils}; \texttt{https://github.com/astropy/specutils}), to which our team aims to contribute SPLAT tools. It is also desireably to expand SPLAT's scientific applications beyond ultracool dwarfs to, e.g., planetary bodies (e.g., asteroidal surface composition modeling), exoplanets (e.g., transmission spectroscopy), and galaxies (e.g., population synthesis modeling); as well as mission-specific tool development (e.g., HST/WFC3; Euclid). We are also developing a web portal for SPLAT tools which will allow users without formal Python/programming backgrounds to explore these data, a task well-suited for K-12 education and citizen science research. 

\section*{Acknowledgements}
This work is supported by the National Aeronautics and Space Administration under Grant No.\ NNX15AI75G.



\label{lastpage}
\end{document}